\documentclass[aps,prl,twocolumn,groupedaddress,showpacs]{revtex4}

\usepackage{graphicx}
\usepackage{amsmath}
\usepackage{amssymb}

\begin{document}


\title{Phase Behavior of Bent-Core Molecules}

\author{Yves Lansac}
\author{Prabal K. Maiti}
\author{Noel A. Clark}
\author{Matthew A. Glaser}
\affiliation{Condensed Matter Laboratory, Department of Physics, and \\
         Ferroelectric Liquid Crystal Materials Research Center, \\
         University of Colorado, Boulder, CO 80309, USA}

\date{}                              


\begin{abstract}                                               

Recently, a new class of smectic liquid crystal phases (SmCP phases)
characterized by the spontaneous formation of macroscopic chiral domains
from achiral bent-core molecules has been discovered.
We have carried out Monte Carlo simulations of a minimal 
hard spherocylinder dimer model to investigate the role of excluded volume 
interations in determining 
the phase behavior of bent-core materials and to probe the molecular 
origins of polar and chiral symmetry breaking. 
We present the phase diagram as a function of pressure or density and dimer
opening angle $\psi$.
With decreasing $\psi$, a transition from a nonpolar to a polar smectic phase
is observed near $\psi = 167^{\circ}$, and the nematic phase becomes 
thermodynamically unstable for $\psi < 135^{\circ}$.
No chiral smectic or biaxial nematic phases were found.

\end{abstract}                                                


\maketitle
%


Molecular chirality plays an important role in the science of liquid crystals 
(LCs), leading to cholesteric LCs \cite{maugin}, blue phases \cite{blue},  
ferroelectric \cite{meyer} and antiferroelectric \cite{fukuda} smectic 
phases, and twist grain boundary phases \cite{renn}.
In all of these examples, chirality is an intrinsic property built into the 
chemical structure of the LC molecules.
Recently, a new class of smectic LC phases (SmCP phases) characterized by 
the spontaneous formation of macroscopic chiral layers from achiral 
molecules has been discovered \cite{niori,link}.
The molecules comprising these phases have `bow' or `banana' shaped cores. 
The resulting phases exhibit two spontaneous symmetry-breaking instabilities: 
polar molecular orientational ordering within the layer plane, and molecular 
tilt, which together produce chiral layers with a handedness that depends
on the direction of the tilt relative to the polar axis.
Very large second order nonlinear optical (NLO) coefficients have been 
measured 
in the ferroelectric state of such materials, bearing some promising 
applications in NLO devices \cite{heppke,araoka}. 

From a theoretical point of view, the relationship of phase behavior
to the specific bent-core molecular shape is of fundamental interest.
In this paper, we investigate a minimal excluded volume model of bent-core
mesogens, focusing on the molecular origin of polar and/or chiral symmetry
breaking. Of particular interest is the coupling between polar and chiral
symmetry breaking. In all bent-core materials studied to date, polar
symmetry breaking is accompanied by chiral symmetry breaking induced by 
molecular tilt. This empirical fact raises the question whether there is
a fundamental connection between polarity and chirality in molecular fluids.
Another empirical observation is that bent-core materials exhibiting
SmCP phases generally do not exhibit nematic phases, although two exceptions
have recently been reported \cite{weissflog1, tschierske}.  One objective
of this study is to establish the molecular shape requirements for the
occurence of the nematic phase in bent-core materials. Finally, we explore
the possibility of {\sl biaxial} nematic ordering in bent-core materials, 
motivated by recent experimental indications \cite{tschierske, kumar}.

Hard core models are particularly appealing due to their simplicity and 
relative ease of
computation, both in simulation and theory.
In particular, hard spherocylinders have been widely studied as simple models 
for conventional LCs \cite{jackson,bolhuis}. This model exhibits rich phase 
behavior including isotropic, nematic, smectic, columnar and solid phases, 
the phase transitions being driven by the competition of two main entropic 
contributions, the orientational entropy favouring the isotropic phase and 
the positional entropy favouring ordered phases, as shown in 
the forties by Onsager in the limit of infinitely thin rods 
\cite{onsager}.

\begin{figure}[!tbp]
\includegraphics[width=3.3in]{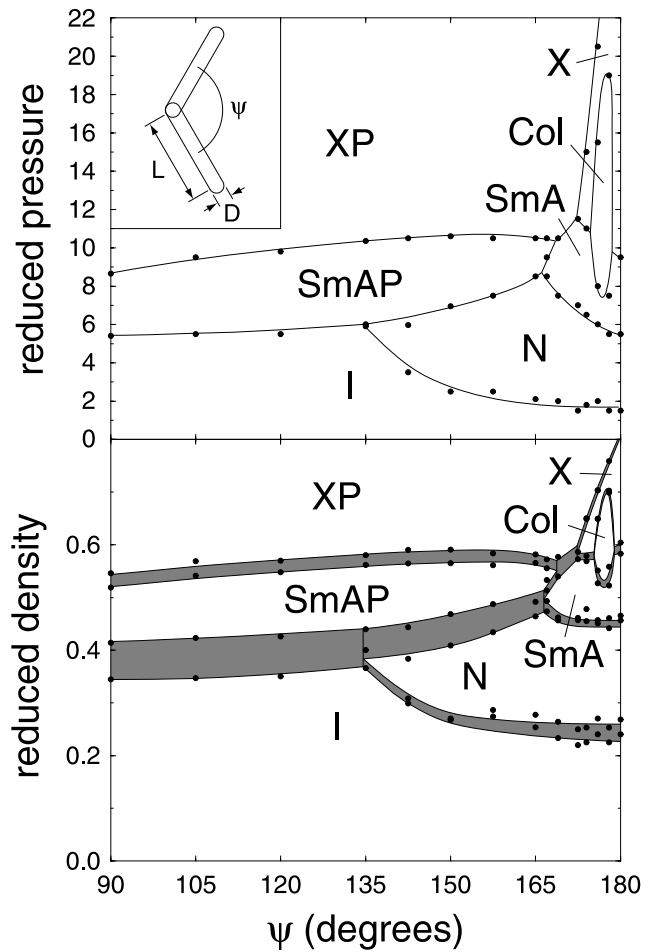}
\caption{ \label{figure:phase}
Phase diagram of spherocylinder dimers (inset)
with aspect ratio $L/D = 5$ as a function of opening angle $\psi$
and reduced pressure $P^{\ast}$ (top) and
reduced density $\rho ^{\ast}$ (bottom). All two-phase regions are 
shaded. The following phases are present: isotropic liquid (I),
nematic (N), polar smectic A (SmAP), smectic A (SmA), columnar (Col),
polar crystal (XP) and crystal (X).
}
\end{figure}

To capture the main characteristics of the collective behavior of
bent-core molecules, we extend the spherocylinder model by introducing 
a hard-core dimer formed by two interdigitated hard-core
spherocylinders sharing one spherical end cap (see inset in
Figure~\ref{figure:phase}).
This is an ideal model system to consider due to the relatively small parameter
space. There are three parameters: two geometrical parameters, namely  
the length-to-breadth ratio $L/D$,
where $L$ and $D$ are respectively the length and the diameter of each 
spherocylinder, and the opening angle $\psi$ between the two spherocylinder
axes; and one thermodynamic parameter, the  
reduced pressure $P^{\ast}$ defined as $P^{\ast} = 
\beta P {v_{o}}$ or, equivalently, the reduced density ${\rho}^{\ast}$ 
defined as ${\rho}^{\ast} = \rho {v_{o}}$. Here, ${v_{o}}$ is the volume
of the equivalent straight hard spherocylinder ($\psi$ = 180$^{\circ}$), 
${v_{o}} = \pi {D^{3}} / 6 + \pi {L'} {D^2} / 4$ with ${L'} = 2 L$.
In all the simulations presented below, we consider a single value for the 
length-to-breadth ratio, namely $L/D = 5$.  This ratio has been chosen in order 
to roughly represent the equivalent geometrical envelope described by a 
realistic bow shaped molecule, with fully extended aliphatic tails.

Using a similar model, Camp {\sl et~al.} \cite{camp} have found a nematic
phase and a smectic A phase of bent-core molecules roughly half as long 
as the ones considered here. 
However, no systematic study of the phase behavior was carried
out. Recently, a polar smectic A phase and a chiral crystal phase have
been observed for a `polybead' model of bent-core molecules with an opening
angle of $\psi = 140^{\circ}$ \cite{selinger}.

\begin{figure}[!tbp]
\includegraphics[width=3.3in]{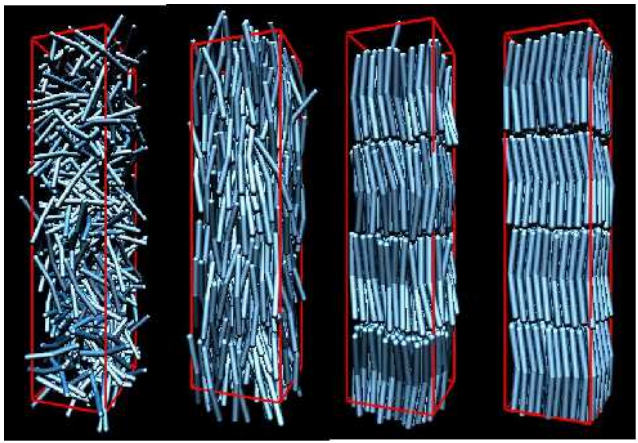}
\caption{ \label{figure:snap}
Final configurations from Monte Carlo simulations of N = 400 
bent-core molecules with opening angle $\psi = 165^{\circ}$ as a function
of pressure. From
left to right: isotropic phase ($P^{\ast} = 1$), nematic phase 
($P^{\ast} = 5$), polar smectic A ($P^{\star} = 10$), and polar crystal 
($P^{\star} = 15$).
}
\end{figure}

To investigate the phase behavior of our model as a function of the
pressure and of the opening angle $\psi$, we perform $NPT$ Monte Carlo (MC)
simulations, with periodic boundary conditions, on a system of $N$ = 400 
bent-core molecules. 
For each opening angle, the system is initially prepared at high pressure
in the crystal phase corresponding to the highest number density 
(antipolar crystal, see Figure~\ref{figure:snap}). The equation of state
$P ^\ast$($\rho ^\ast$) is determined by measuring the density $\rho ^\ast$
as the pressure is decreased incrementally from the crystalline state.
At each state point ($\psi$, $P^\ast$), $2\times 10^5$ MC cycles are used 
for equilibration and $1 \times 10^6$ MC cycles are used for production 
of the results.

The location of the phase boundaries is determined from the equation of state
$P ^\ast$($\rho ^\ast$), and the nature of the coexisting phases is 
investigated through the computation of various order parameters.
In-layer crystalline order is probed with the translational order parameters 
${{\rho}_{\bf G}}_{k}$ defined as 
${{\rho}_{\bf G}}_{k} = \frac{1}{M} \sum_{j = 1}^{M} \exp ({i {\bf G}_{k}
\cdot {\bf r}_{j}})$ where ${\bf G}_{1}$, ${\bf G}_{2}$ and ${\bf G}_{3}$
are the reciprocal basis vector of a deformed hexagonal lattice,
${\bf r}_{j}$ is the position of the center
of mass of the molecule $j$, and $M$ is the number of molecules in a 
given layer. 
The smectic order is measured by the layer translational order 
parameter ${\rho}_{\parallel}$ defined as
${\rho}_{\parallel} = \frac{1}{N} \sum_{j = 1}^{N} \exp ({i {\bf G}_{\parallel}
\cdot {\bf r}_{j}})$ with 
${\bf G}_{\parallel} = \frac{2 \pi}{d} \; \hat{{\bf z}}$, where $d$ is the 
layer spacing and $\hat{{\bf z}}$ is the layer normal. 
Polar orientational order is detected using the polar order parameter 
${{\bf m}}$ 
defined as ${{\bf m}} = \frac{1}{M} \sum_{j=1}^{M} \hat{{\bf m}}_{j}$ where
$\hat{{\bf m}}_{j}$ is a unit vector contained in the plane of the
molecule and passing through the apex of the molecule \cite{note}. 
Finally, molecular orientational order
is determined from the largest eigenvalue of the second-rank tensorial 
orientational order parameter $Q_{\alpha \beta}$, defined as  
$Q_{\alpha \beta} = \frac{1}{N} \sum_{j=1}^{N} \left({ \frac{3}{2}
n_{i_{\alpha}} n_{j_{\beta}} - \frac{1}{2} {\delta}_{\alpha \beta} }\right)$,
where $\alpha, \beta = x,y,z$, 
and $\hat{{\bf n}}_{j}$ is a unit vector parallel to the molecular 
end-to-end vector for molecule $j$.

The ($\psi, P^{\ast}$) and ($\psi, {\rho}^{\ast}$) phase diagrams are
presented in Figure~\ref{figure:phase}.
Rich phase behavior is found, with isotropic (I), nematic (N), 
smectic A (SmA), polar smectic A (SmAP), columnar (Col), crystal (X) and 
polar crystal (XP) phases. Configurations from the isotropic, nematic, 
polar smectic and polar crystalline phases are shown in 
Figure~\ref{figure:snap}, for an opening angle of $\psi = 165^{\circ}$. 

The nematic phase is stable  for opening angles larger than 
$\psi$ = 135$^{\circ}$. 
With decreasing opening angle, the region of stability of the nematic
phase decreases, vanishing for opening
angles smaller than $\sim 135 ^{\circ}$, leading to an (I, N, SmAP) triple
point near $\psi = 135 ^{\circ}$.
It is interesting to note that for dimers half as long as ours, 
small opening angles seem to destabilize
the smectic phase rather than the nematic phase \cite{camp}.

The vast majority of bent-core materials 
exhibiting liquid crystal behavior, and in particular SmCP phases, have 
an opening angle between $120 ^{\circ}$ and $135 ^{\circ}$, and do 
not exhibit any nematic phase.  
Quite interestingly, two classes of bent-core compounds 
having an opening angle between $134 ^{\circ}$ and $148 ^{\circ}$ 
\cite{weissflog1,dingemans} exhibit both smectic and nematic phases.
These observations, in good qualitative agreement with the 
predictions of the model, tend to confirm the hypothesis that excluded volume
interactions play a central role in the behavior of such materials. 

The existence of a biaxial nematic phase remains an elusive possibility in
thermotropic LCs. Due to their bent-core geometry, banana molecules
are good candidates to investigate such phenomena. 
Recent experiments suggest that the nematic phase exhibited by two
classes of bent-core material might be biaxial \cite{tschierske, kumar}. 
However, the nematic phase presented by our hard-core model does not exhibit 
any biaxiality.
It is likely that the reported biaxiality is due to more subtle
interactions and/or to the presence of the flexible tails. 
A biaxial nematic phase has been reported for hard
spherocylinder dimers twice as long as the ones used in the present study,
but no transition from/to an isotropic liquid has been reported \cite{camp}.
Using a simplification of the Onsager second-virial treatment and bifurcation
analysis, Teixeira {\sl et al.} have found a biaxial nematic phase in the 
limit of very long bent-core molecules \cite{teixeira}. 

Because straight spherocylinders do not exhibit any polar smectic ordering, 
it is expected that our model should exhibit a transition from nonpolar 
smectic (SmA) to polar smectic (SmAP). This transition occurs for an opening 
angle between $167 ^{\circ}$ and $169 ^{\circ}$, and is associated with 
two triple points, a (SmA, SmAP, N) triple point near 
$\psi = 167^{\circ}$ and a
(SmA, SmAP, XP) triple point near $\psi = 169^{\circ}$. Very recently,
the first group of bent-core molecules exhibiting both SmCP and a
SmA (as well as SmC and nematic phases in one case) have been synthesized 
\cite{weissflog1,weissflog2}. The opening angles measured by NMR techniques
for the region of appearence of the smectic A remains in the range 
of 132$^{\circ}$ to 145$^{\circ}$ but no polar order has been detected 
\cite{huang}, lowering the upper limit of stability of a polar smectic phase
with respect to our predictions.

\begin{figure}[!tbp]
\includegraphics[width=2.5in]{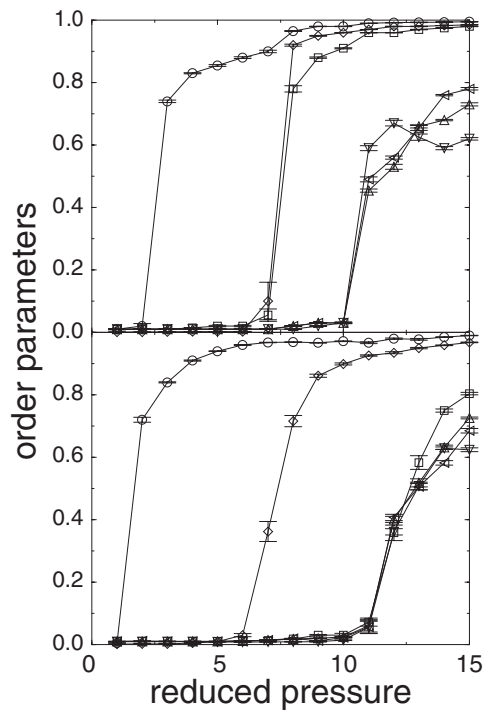}
\caption{ \label{figure:order}
Evolution of the squared-magnitude of order parameters as a function of reduced
pressure for an opening angles of $\psi = 157.5 ^\circ$ (top) and 
$\psi = 172.5 ^\circ$ (bottom) showing, respectively
the phase sequences XP--SmAP--N--I and XP--SmA--N--I as a function 
of decreasing pressure.
The following order parameters are plotted: ($\vartriangle$, $\triangledown$,
$\vartriangleleft$)
solid-liquid order parameters ${\rho}_{1}$, ${\rho}_{2}$, ${\rho}_{3}$;
($\square$) polar order parameter
${\bf m}$; ({\Large$\diamond$}) smectic order
parameter $\rho_{\parallel}$; ({\Large$\circ$}) the largest eigenvalue
of the nematic order parameter $Q_{\alpha \beta}$.
}
\end{figure}

Figure~\ref{figure:order} displays the evolution of the order parameters
for opening angles $\psi = 157.5^{\circ}$ and $\psi = 172.5^{\circ}$. 
In the former case, a polar smectic phase 
characterized by a high value of both the smectic and polar order parameters
is present while in the later case the appearance of a smectic phase from
the higher density crystal phase is accompanied by 
a jump to zero of both the crystal order parameters and the  
polar order parameter.  
The smectic phases (both polar and nonpolar) do not exhibit any 
tilt. 
Neverthless, our results tend to confirm the hypothesis that the polar ordering
is related to the distinctive bent-core shape of the molecules but is not
strongly related to the molecular tilt. Thus, polarity does not imply
chirality. 

A transition between a polar crystalline phase and a 
narrow  nonpolar crystalline phase (i.e a rotator phase) is also present.
This rotator phase is stable for opening angles larger than $\psi =
$ 172.5$^{\circ}$,
and is characterized by a (SmA, X, XP) triple point around 
$\psi =$ 172.5$^{\circ}$.  
Quite interestingly, the rotator phase competes with a columnar phase for
opening angles larger than $\psi =$ 174$^{\circ}$ and smaller than 
$\psi =$ 180$^{\circ}$.
This narrow columnar phase is characterized by a significant crystal 
order parameters but a negligible magnitude of the 
smectic order parameter. Since no evidence of such a phase was found for
straight spherocylinders, a slightly bent molecular shape seems to favor the
appearence and the stabilization of a columnar phase.
Due to the rather unexpected appearence of the columnar phase, we performed
additional studies of this region of the phase diagram using helical periodic
boundary conditions \cite{allen} for N = 400 
and a direct `quench' from a crystalline state to the middle of the columnar 
phase for $\psi = 176^{\circ}$ with helical periodic conditions for a 
larger system (N = 1600). A columnar phase was observed in both studies.
However, we feel that free energy computations 
are needed in order to assess the relative thermodynamic stability of 
the columnar phase, the nonpolar smectic phase, and the rotator phase.

Insights into the shape of the phase boundaries can be gain by supposing,
to a first approximation, that the partition function of the system can be
decomposed into a product of a positional and orientational contributions, 
in which case the entropy is the sum of an orientational entropy 
and a translational entropy.
Competition between different forms of entropy 
determines the stability of a given phase at a given density. In the limit
of straight spherocylinders, the isotropic--nematic phase transition occurs
when the gain in positional entropy $S^{\rm pos}$ exceeds the loss of 
orientational entropy $S^{\rm orient}$ \cite{onsager}. 
A nematic--smectic phase transition 
occurs when the gain in translational entropy perpendicular to the long 
molecular axis $S^{\rm pos}_{\perp}$ exceeds the loss of positional
entropy parallel to the long molecular axis $S^{\rm pos}_{\parallel}$, leading 
to the formation of a stack of two-dimensional liquid layers. 
Similar reasoning can be applied to bent-core molecules: in the range 
$134 ^{\circ} < \psi < 180 ^{\circ}$, the isotropic phase is more favourable 
at smaller 
opening angles. As the cores become more bent,  
the gain in positional entropy associated with nematic ordering is reduced. 
The nematic phase range is then reduced, eventually disappearing for 
$\psi < 134^{\circ}$. 
The shape of the nematic--SmAP boundary (i.e., for $134^{\circ} < \psi < 
167^{\circ}$)
can be qualitatively understood in the same way by noticing that the 
positional entropy parallel to the long molecular axis 
$S^{\rm pos}_{\parallel}$ is larger for larger opening angles than for 
smaller ones, stabilizing the nematic phase for larger
opening angles. This trend is reversed for the nematic--SmA transition (i.e.,
for $\psi > 167 ^{\circ}$) because the absence of polar order
leads to jamming, reducing
the translational entropy perpendicular to the long molecular axis 
$S^{\rm trans}_{\perp}$ for decreasing opening angles. This effect
is responsible for the enhanced stability of the nematic phase for decreasing
opening angles. 

The spherocylinder dimer model exhibits rich phase behavior, 
including polar and nonpolar crystal, columnar, polar and nonpolar smectic,
nematic and isotropic phases.
In particular the existence and range of stability of the nematic phase
is in good agreement with the behavior of the new class of bent-core
molecules, while the stability of the nonpolar smectic
A phase is in qualitative agreement with experiments.
In addition our model predicts the existence and stability of a polar
smectic A phase for $\psi < 167^{\circ}$. Such phases remain to be found.
No tilted phases are exhibited by the model, and a simple extension of the 
present model, taking into account the important
steric role played by the flexible LC tails in the formation of tilted smectic
phases, is currently under investigation.
Our model indicates that there is no intrinsic coupling between polar 
symmetry breaking and chiral symmetry breaking, and that the later is not 
directly related to the bent-core shape of the molecules. 

\acknowledgments

This work was supported by NSF MRSEC Grant DMR 98-09555.




\begin{references}
\bibitem{note} No strong polar coupling between layers
has been found in our simulation. In the remainder of this article, we take
polar ordering to mean polar ordering within a layer.
\end{references}
\end{document}